\documentclass[%
twocolumn,
 fleqn,usenatbib,
 floatfix,nolongbibliography,aps,
author-numerical%
]{revtex4-2}

\pdfoutput=1
\usepackage[normalem]{ulem} %this is for use of deleted font
\usepackage[T1]{fontenc}
\usepackage [latin1]{inputenc}
\DeclareRobustCommand{\VAN}[3]{#2}
\let\VANthebibliography\thebibliography
\def\thebibliography{\DeclareRobustCommand{\VAN}[3]{##3}\VANthebibliography}

\usepackage{newtxtext,newtxmath}
\usepackage{tikz,xcolor,hyperref}
\usepackage{graphicx,url,caption,subcaption}% Include figure files
\usepackage{dcolumn}% Align table columns on decimal point
\usepackage{bm}% bold math
\draft % marks overfull lines with a black rule on the right

\begin{document}

%\preprint{AIP/123-QED}

% Make Orcid icon
\definecolor{lime}{HTML}{A6CE39}
\DeclareRobustCommand{\orcidicon}{%
	\begin{tikzpicture}
	\draw[lime, fill=lime] (0,0) 
	circle [radius=0.16] 
	node[white] {{\fontfamily{qag}\selectfont \tiny ID}};
	\draw[white, fill=white] (-0.0625,0.095) 
	circle [radius=0.007];
	\end{tikzpicture}
	\hspace{-2mm}
}

\foreach \x in {A, ..., Z}{%
	\expandafter\xdef\csname orcid\x\endcsname{\noexpand\href{https://orcid.org/\csname orcidauthor\x\endcsname}{\noexpand\orcidicon}}
}

% Define the ORCID iD command for each author separately. Here done for two authors.
\newcommand{\orcidauthorA}{0000-0001-9180-4773}
\newcommand{\orcidauthorB}{0000-0002-6886-1398}
%%%%%%%%%%%%%%%%%%%%%%%%%%%%%%%%%%%%%%%%%%%%%%%%%%

\title[2D EM waves in plasma-metamaterials]{Two dimensional modelling of the interaction between electromagnetic waves and plasma-metamaterial composite structures using the particle-in-cell method}% Force line breaks with \\

\author{D. Tsiklauri\orcidA{}}
 \email{D.Tsiklauri@salford.ac.uk}
\affiliation{Joule Physics Laboratory,
School of Science, Engineering and Environment, 
University of Salford,
Manchester, M5 4WT, 
United Kingdom}
\author{I. Morrison\orcidB{}}
\affiliation{Joule Physics Laboratory,
School of Science, Engineering and Environment, 
University of Salford,
Manchester, M5 4WT, 
United Kingdom}

\date{\today}% It is always \today, today,
             %  but any date may be explicitly specified

\begin{abstract}
In this work we (i) extend previous 1D studies
of electromagnetic (EM) wave propagation in an over-dense 
plasma-metamaterial composite into two spatial dimensions and
(ii) study blocking of EM waves by the composite
2D structures (barriers).
Such barriers are formed when metamaterial spatially
co-exists with a plasma density depletion
in a form of a slab 
 or two-dimensional density rectangular 
depletions (DRDs). This is analogous to EM wave
trapping by preformed density cavities 
in near-critical density plasmas, studied before.
We find that plasma-metamaterial composite 
allows to block EM waves by both slab and DRD configurations,
thus forming a standing wave at the edge of an opaque region.
The standing wave subsequently damps which offers
applications such as heat deposition or 
substrate materials (micro)machining depending on EM wave intensity.
The established results may find  
future applications such as: more efficient plasma vapour
deposition, controlling EM wave propagation 
(EM wave blocking) in invisibility cloaks and alike.
The EM wave blocking conditions are elucidated by a set of
particle-in-cell (PIC) numerical simulations.
%168 words
\end{abstract}

%\keywords{instabilities -- magnetic fields -- plasmas -- 
%Sun: heliosphere -- ISM: magnetic fields}%Use showkeys class option if keyword
                              %display desired
\maketitle

\section{Introduction}
Ref. \cite{iwai2017} gives a good background and lays a foundation to the 
considered research. It is well-known that
industrial applications such as 
{physical vapor deposition (PVD)}, 
material surface treatment and many others use plasma as an
agent or a means of delivery \cite{lieberman1994}. 
This technology bridges the gap between
conventional thermal spray and vapor deposition and
provides a variety of coating  microstructures
composed of vapor, liquid, and solid deposition units.

A common method of creation of plasma in these applications is gas irradiation by microwaves. Microwaves
are a form of electromagnetic 
radiation with frequency range between 1 and 100 GHz (wavelengths
between 300 and 3 mm). When a microwave enters a vessel with the plasma discharge, and the electron
density $n_e$ is larger than the cut-off density for the
incident microwave, the latter cannot propagate
through plasma and is confined to its surface \cite{sugai1998}.
In microwaves, the surface wave, with a typical
width of electron inertial length $c/\omega_{\rm pe}$, sustains
the over-dense plasma, but the width of the
generated plasma is narrow. 
{Here the term over-dense plasma has the 
following meaning: 
Plasma is only transparent to EM wave with a frequency 
$\omega_0$ larger than the electron 
plasma frequency 
$\omega_{\rm pe} = \sqrt{n_e e^2/(m_e \varepsilon_0)}$, where
$n_e$ is the electron number density.
In such case permittivity $\varepsilon_p =1-\omega_{\rm pe}^2 /\omega_0^2$
is a positive, real number. In the opposite case
when $\omega_0<\omega_{\rm pe}$ the permittivity $\varepsilon_p$ becomes
negative and plasma is opaque to such EM wave.}
In order to increase
penetration of microwaves into the over-dense
plasma, DC magnets or magnetic field coils are
used. The magnetized plasma has the
propagation band under the electron cyclotron
frequency, and this enables to optimize this
frequency by the strength of the external
magnetic field. This propagation mode is the
whistler mode, and a wavenumber becomes
infinity when the incident wave frequency
approaches the electron cyclotron frequency. The
electron cyclotron resonance (ECR) heating is
useful for the efficient power injection into the
overdense plasma, but it is difficult to prepare a
large-scale DC magnet for the plasma generation
systems. Metamaterials are artificial composites
made of unit patterns whose size is much smaller
than the wavelength of the corresponding waves.
Metamaterials
 give  the  extraordinary
propagation of EM waves and it
cannot occur in natural materials. Refractive
index $N$ becomes negative in metamaterials
because both permittivity $\varepsilon$ and permeability $\mu$
become negative \cite{veselago1968}. 
The experimental verification was
performed with the combination of the split-ring
resonators (SRRs) as negative-$\mu$ metamaterials \cite{pendry1999}.
Theoretical and numerical simulation studies of wave propagation in
negative $N$ are numerous \cite{ziolkowski2001,foteinopoulou2003,agranovich2004}. 

Refs. \cite{agranovich2004,shadrivov2006} reported the generation of the second
harmonic waves in negative $N$ with quadratic
nonlinear response. Also the
nonlinear frequency conversion process was
demonstrated to be analogous to non-linear
optics \cite{boyd2008}. 
Also, experimental studies were
performed with the combination of the arrays of metal wires as a negative-$\varepsilon$ material \cite{shelby2001,houck2003,parazzoli2003}.
The array of metal wires has a cut-off frequency 
and acquires negative $\varepsilon$ under the cut-off frequency as in the
over-dense plasma \cite{rotman1962}. 
A composite  of an overdense plasma and a negative-$\mu$  metamaterial has
been investigated experimentally 
\cite{nakamura2014,iwai2015a,iwai2015b} and by means of numerical simulations \cite{sakai2011,kourtzanidis2016}. 
It was demonstrated that 
 that refractive index $N$ was
negative, the second harmonic wave was present in Ref. \cite{iwai2015a}, and SRRs were electrically connected with plasma
at a microscopic level \cite{iwai2015b}, wave propagation was not clearly confirmed in the composite of an overdense plasma
and a negative-$\mu$ metamaterial.
Ref.\cite{iwai2017} conducted 1D electromagnetic PIC simulations 
of EM wave propagation in the composite of
overdense plasma and the negative-$\mu$ metamaterial. 
In their simulation EM
waves enter the plasma-metamaterial composite, 
and propagate in it with a negative phase
velocity and a positive group velocity. This way,
Ref.\cite{iwai2017} confirmed that the plasma-metamaterial composite has a
negative refractive index.
Ref. \cite{iwai2020} gives an uptodate background to the experimental
implementation of the considered research.
{In particular \citet{iwai2020} conducted
measurements of properties of transmitted microwaves and
plasma parameters in a composite of double-split-ring resonators (DSRRs),
which make magnetic permeability negative. In their experiments 
plasma was created using a 2.45 GHz microwave. 
They launched the microwaves with power in the range of
$100-200$ W, modulated by a pulse wave at a low frequency and a low duty ratio to detect time evolutions of plasma parameters. This way, 
\citet{iwai2020} performed the measurements 
at different positions with varying distance 
from the supporting plate of the DSRRs, 
and established the enhanced wave transmission 
in the composite with non-uniform profiles 
of electron temperature. 
They confirmed that the non-uniformity is due to
the magnetic resonant behaviour 
of the DSRRs at the microscopic level, 
while their macroscopic manifestation is making
the permeability negative, enhancing wave energy 
transfer inside the composite.}

Ref.\cite{sakai2011} performed FDTD simulations and 
reported the wave propagation with a negative $N$ in
the array of plasma columns under a negative-$\mu$ state. 
They assume that a propagating microwave
contributes to simultaneous plasma generation by its electric field $E$ that deforms $\varepsilon$ and refractive index $N$ in
the metamaterial; this is a field-dependent metamaterial, in which a propagating wave conversely suffers
from the changes of $\varepsilon$ and/or $N$, where $E$, $\varepsilon$ , and $N$ settle self-consistently. Such a metamaterial may exhibit
nonlinear features due to its field-dependent properties.
Ref.\cite{kourtzanidis2016} studied the local
electromagnetic fields around an SRR when a
plasma exists near the SRR. Since both reports
used the FDTD solver to simulate EM wave
coupling to plasmas i.e. coupling between high-frequency EM waves and the plasma occurs mainly
via the electron current density (ions are heavy
enough not to respond in EM field variations).
Hence in both Ref.\cite{sakai2011} and 
Ref.\cite{kourtzanidis2016} the kinetic effects
such as self-consistent EM fields from collection
of billions of particles in plasma were not
considered. These effects must be included in the
case when a high-power electromagnetic wave enters
the overdense plasma under a negative-$\mu$ state.

Although, Ref.\cite{iwai2017} conducted fully kinetic
 electromagnetic PIC simulations 
of EM wave propagation in the composite of
overdense plasma and the negative-$\mu$ metamaterial,
they used PIC code modified from KEMPO1 which is
\textit{spatially 1 dimensional (1D)}. 
There is a clear need to extend model of Ref. \cite{iwai2017}
 beyond simple 1D case. This
has very strong industrial applications in plasma processing. i.e. use of negative-$\mu$ metamaterial (the plasma-metamaterial composite) 
will enable to increase plasma density 
which is crucial for plasma processing
efficiency. It can be also related to modern applications such as ion thrusters and plasma based acceleration,
intense X-Ray sources etc. 
Typical microwaves used today have frequencies in the 
few GHz range and the
cut-off density that corresponds to the cut-off 
frequency $f_{\rm pe}= \omega_{\rm pe} /2 \pi$ is few 10$^{16}$ m$^{-3}$. 
If this density can be
increased then plasma processing will become more efficient, 
as the reaction rate is usually proportional to the
product of number densities of the reactant species. 
Thus, in this work propose to extend previous
1D results \cite{iwai2017} into two spatial dimensions and use
the modified 2D  (and in the future 3D) 
EPOCH PIC code \cite{arber2015}. 
The key novelty is that we propose to use much more general,
fully kinetic regime using EPOCH code, while all previous research 
\cite{sakai2011,kourtzanidis2016} used FDTD method to simulate
plasmas, hence kinetic effects of a plasma and non-linearity were not considered. These effects must be
included when
a high-power EM wave enters the overdense plasma under a negative-$\mu$
state. 

The second input and motivation for the present study comes
from research related to the 
trapping of EM waves 
in preformed density cavities in near-critical density plasmas 
\cite{yu1978,wang2009,luan2012,zhu2012,luan2013}.
In a more broader context, the present research is about
an ability to control and manipulate EM radiation by means of combining
the presence of plasma and metamaterials.
In particular Ref. \cite{luan2013} studied
EM wave trapping in cavities in near-critical density plasmas using 2D particle-in-cell (PIC) simulation. They find that in plasma,
 laser's ponderomotive force can create  a vacuum cavity bounded by a thin overcritical-density wall. The EM waves are self-consistently trapped as a half-cycle electromagnetic wave in the form of an oscillon-caviton structure. Further, the trapped EM wave is slowly depleted through interaction with the cavity wall. Ref. \cite{luan2013} also studied a situation when a
 near-critical density plasma contains a \textit{preformed} density
 cavity and found that
 EM wave becomes trapped, forming a standing wave. 
They find that the trapped light is characterized 
as multi-peak structure and that the overdense plasma layer formed 
around the self-generated and preformed 
cavities that is induced by the laser ponderomotive force is the natural reason for EM wave trapping.
Ref. \cite{luan2013} also discusses possible 
applications of the EM wave trapping in 
preformed and self-consistently  generated by EM wave's 
ponderomotive force density cavities in near-critical density plasmas.

We close the introduction by stating 
the aims of this work that are two-fold: 
(i) to extend previous one dimensional studies 
of electromagnetic (EM) wave propagation on an over-dense 
plasma-metamaterial \cite{iwai2017} 
composite into two spatial dimensions and
(ii) to study blocking of EM waves by the composite
barriers in analogy with their 
trapping in preformed density cavities in near-critical density plasmas
\cite{luan2013}.
We therefore propose to use modified 2D EPOCH code,
as described in section 2. 

\section{The model}
We modify the existing state-of-the-art, fully kinetic,
explicit, electromagnetic,
2D, particle-in-cell (PIC) code called EPOCH \cite{arber2015}.
 The modification is by adding the effect of negative-$\mu$ metamaterial (the plasma-metamaterial composite) via adding the 
 magnetic current term, $\mathbf{J_m}$, 
 to the relevant Maxwell's equation as following:
\begin{equation}
\nabla \times \mathbf{E}=-\mathbf{J_m}-\frac{\partial \mathbf{B}}
{\partial t},
\label{eq1}
\end{equation}
\begin{equation}
\frac{\partial \mathbf{J_m}}
{\partial t}=\omega_{\rm m}^2 \mathbf{B},
\label{eq2}
\end{equation}
where where $\omega_{\rm m}$ is the magnetic resonance frequency.
{It does not have 
any associated mathematical expression like the electron 
plasma frequency has $\omega_{\rm pe} = \sqrt{n_e e^2/(m_e \varepsilon_0)}$.
$\omega_{\rm m}$ is just a constant
that quantifies the effect of presence metamaterial
in the numerical simulation.
In the experiments such as \citet{iwai2020} $\omega_{\rm m}$
is prescribed by the material properties and physical dimensions of DSRRs}. 
The modification of EPOCH code is
insertion of $\mathbf{J_m}$. The metamaterial effect, based on $\mathbf{J_m}$,
does not affect both electric field,
 $\mathbf{E}$, and plasma in this simulation. $\mathbf{J_m}$ is
determined only by $\mathbf{B}$ and the constant value of 
$\omega_{\rm m}$. 
{According to
Ref.\cite{iwai2017} such approach has 
proven to be very effective in modeling of
bulk effects of negative-$\mu$ metamaterial.
We would to remark  that  to model the response 
 of a magnetic negative (MNG) material 
we also use, as in \cite{iwai2017},
 the Drude model, and as for 
 the response of the epsilon negative (ENG) permittivity of plasma. 
 However,  MNG 
 materials are usually obtained using resonant substructures such as 
 split-ring resonators (SRR) which are generally 
 not accurately modeled with the 
 Drude model. In particular, perhaps some other, a more
 accurate model, is needed to better describe 
 the narrow band behavior of the MNG 
 material in the PIC simulation. This can be a 
 subject to future investigation.}
The plane, linearly-polarized EM
wave is excited by the external current
$\mathbf{J_s}$ at $x=1\Delta$, the first grid point
of the simulation domain, where $\Delta=r_D=
\sqrt{\varepsilon_0 k_B T_e/(n_e q_e^2)}$ is the Debye length.
$\mathbf{J_s}$ oscillates in
vertical
 $z$-direction 
$\mathbf{J_s} = \hat z J_z =J_0 \sin( \omega_0 t )$.
 EM wave travels in $x$-direction and
 has components $E_z$ and $B_y$ as it
travels initially in free-space. The plasma-metamaterial composite exists in the region where $12.8 < x <25.6$, see Figure \ref{fig2}(a),
 that we  call \textit{plasma region}. 
{The sketch of the numerical simulation is very similar
to Fig.1 from \citet{iwai2017}, so it is not duplicated here.} 
Note that spatial dimensions are normalized by the EM wave inertial length
$c/\omega_0$.
 When EM wave enters the plasma region it excites a different current $\mathbf{J_p}$ (subscript p stands for plasma),
  which is prescribed
by the following equations:
\begin{equation}
\nabla \times \mathbf{B}=\mu_0
\mathbf{J_p}+\mu_0\varepsilon_0\frac{\partial \mathbf{E}}
{\partial t},
\label{eq3}
\end{equation}
\begin{equation}
\frac{d \left(\gamma v_\alpha\right)}
{d t}=\frac{q_\alpha}{m_\alpha}\left(\mathbf{E}+\mathbf{v_\alpha}
\times \mathbf{B} \right),
\label{eq4}
\end{equation}
where $\mathbf{J_p}$ in equation (\ref{eq3})  is different from the 
initial $\mathbf{J_s}$ because when EM wave enters plasma-metamaterial composite,
electrons and ions start to oscillate and $\mathbf{J_p}$ is now prescribed by dielectric response of plasma and velocity of
oscillating plasma particles of species $\alpha$, $\mathbf{v_\alpha}$. 
EPOCH is a relativistic code therefore, 
note that in Eq.(\ref{eq4}),  $\gamma$ is the Lorentz $\gamma$-factor.
In our case it is very close to unity.
Note that when EM wave amplitude and/or plasma number density are small,
plasma has the simple dispersive back-reaction to EM wave's $\mathbf{E}$ 
as $\partial \mathbf{J_p}/\partial t \simeq \omega_{\rm pe}^2 \mathbf{E}$ where $\omega_{\rm pe}$ is the plasma
frequency. When metamaterial is present its effect is felt by resonant interaction of EM wave's magnetic
field $\mathbf{B}$. The effect of magnetic resonance can be described by appearance of another frequency $\omega_{\rm m}$ , which
appears in equation (\ref{eq2}). 
Note that in equation (\ref{eq2}), the term
$\partial \mathbf{J_m}/\partial t = 
\omega_{\rm m}^2 \mathbf{B}$
is mathematically similar to $\partial \mathbf{J_p}/\partial t \simeq 
\omega_{\rm pe}^2 \mathbf{E}$. 
This effectively means that
presence of metamaterial is manifest by additional current $\mathbf{J_m}$. In this formulation, the metamaterial  
does not affect $\mathbf{E}$ and plasma in this simulation. $\mathbf{J_m}$ is determined only by $\mathbf{B}$ and the constant $\omega_{\rm m}$. 
Note that
plasma current $\mathbf{J_p}$ is prescribed by nonlinear and kinetic motions of the particles and consists of the $x$- and $z$-
axis components. The magnetic current density $\mathbf{J_m}$  
is parallel to $\mathbf{B}$  and consists of the y-axis component.
Therefore effective permittivity of the plasma $\varepsilon_p$ and 
the effective permeability of the metamaterial $\mu_m$ are
expressed as $\varepsilon_p =1-\omega_{\rm pe}^2 /\omega_0^2$
 and $\mu_m =1-\omega_{\rm m}^2 /\omega_0^2$. This implies that is 
 EM wave's frequency $\omega_0$ is smaller than
both $\omega_{\rm pe}$ and  $\omega_{\rm m}$ ($\omega_0 < \omega_{\rm pe}$
 and $\omega_0< \omega_{\rm m}$) thus both $\varepsilon_p$ and 
$\mu_m$ are negative. Therefore, refractive index 
$N=\sqrt{\varepsilon_p \mu_m}=i^2\sqrt{|\varepsilon_p||\mu_m|}
=-\sqrt{|\varepsilon_p||\mu_m|}$
becomes negative with all of the above described consequences
such a EM wave's negative phase speed propagation in an over-dense
plasma-metamaterial composite region \cite{iwai2017}.

The numerical recipe how to add the additional term responsible for
the presence of the metamaterials is given in appendix A.
Here we focus on the details of the numerical simulation set up.
{We use periodic boundary conditions, 
both in $x$- and $y$-directions,
throughout this work, which numerically are the most precise of
all available boundary conditions. 
In order to avoid the spurious reflections, and to conserve 
the total energy
in the simulation,
we do not to use open or absorbing boundary 
conditions in EPOCH so that to remove the wave 
propagating towards negative $x$ and thus reduce the computational domain.
Our choice of periodic boundary conditions} means that the solution that is
generated at $x=1 \Delta$, 
the first grid cell, that travels to the
left, i.e. negative $x$-direction, and re-enters the simulation
domain from the far right side of the domain $x=x_{\rm max}$,
should not collide with the main studied numerical solution
that travels in positive $x$-direction.
This is achieved by the setting $x=x_{\rm max}=8192 \Delta$.
It can be seen from Fig.\ref{fig1}(a)
that the left- and right- going numerical solutions never collide. 
The above defined notation for  the Debye length
can be also written as $\Delta=\lambda_D=
v_{\rm th,e}/\omega_{\rm pe}=
\sqrt{k_B T_{\rm e,max}/m_{\rm e}}/\omega_{\rm pe}$.
Maximal temperature is set as $T_{\rm e, max}=m_e(0.0375c)^2/k_B$.
We use the term  maximal temperature
because it actually varies across x-axis.
This is because as in Ref.\cite{tsiklauri2024} we keep
pressure balance by making $p_{\rm e,i}=n_{\rm e,i}(x)k T_{\rm e,i}(x)=
const$, i.e. $T_{\rm e,i}(x)=T_{\rm e,i, max}[n_0/n_{\rm e,i}(x)]
\propto 1/n_{\rm e,i}(x)$.
This ensures that the total initial 
pressure balance $p_{\rm e,i}=const$ is fulfilled.
$n_0$ is the background plasma number density,
which is the same both for electrons and protons.
The proton to electron mass ratio is set to the realistic
value of 1836.
The number densities $n_{\rm e,i}$ and 
grid sizes are varied in different numerical runs 
as specified in Table \ref{t1}.

\begin{table}
\captionsetup{justification=raggedright,
singlelinecheck=false}
\caption{Table of numerical runs considered.
See text discussing Fig.\ref{fig3} for
the explanation of the notation used.}
\centering
\begin{tabular}{lcccc} 
\hline
Case &  $n_{\rm e,i}(x)$ & Grid $n_x\times n_y$ & Metamat. & $\rm F$ \\
\hline
1a & $10^{-2}n_0$ & $8192 \times 3$ & N   & - \\
1b & Eq.(\ref{eq5})     & $8192 \times 3$ & N & - \\
1c & $10^{-2}n_0$ & $8192 \times 3$ & Y & -\\
1d & Eq.(\ref{eq5})     & $8192 \times 3$ & Y & - \\ 
2  & Eq.(\ref{eq5})     & $8192 \times 128$ & Y & -\\
3  & Eq.(\ref{eq7})     & $8192 \times 128$ & Y & 0.20 \\
4  & Eq.(\ref{eq7})     & $8192 \times 128$ & Y & 0.99\\
5  & Eq.(\ref{eq8})     & $8192 \times 128$ & Y & 0.20\\
6  & Eq.(\ref{eq8})     & $8192 \times 128$ & Y & 0.99\\

\hline
\end{tabular}
\label{t1}
\end{table}

In EPOCH code physical quantities are in
SI units, so we fix $n_0=10^{15}$ particles per m$^{-3}$
typical of many laboratory (and accidentally astrophysical plasmas too).
In cases 1a or 1c electron and ion number densities are set
to $10^{-2}n_0$ particles per m$^{-3}$.
The factor $10^{-2}$, while drops number density to nearly zero, 
stops EPOCH from slowing down for numerical reasons. 
So effectively the cases with the constant values of $10^{-2}n_0$ 
or other spatial locations, when number density is varying,
can be referred to as 'no plasma', i.e. effectively a vacuum.
 
We use 100 particles per cell for each species, so 
in total, in the large runs, we have 
$2\times100\times8192\times128=2.1\times10^8$
particles. One numerical run takes circa 
2 days, 4 hours and 20 minutes
on 8 processor cores of the 
40-core Intel(R) Xeon(R) CPU E5-2630, 2.20GHz Linux server.
We checked numerical convergence of the
presented results. Twice coarser grid and twice less number of
particles yields numerical results which look the same at 
plotting accuracy i.e. 1 $\%$.
The length is normalized on $c/\omega_0$.
The end simulation time is set $t_{\rm end}=100/\omega_0$. 
Note that electron inertial length,
$c/\omega_{\rm pe}$, is resolved with 27 grid points, i.e.
$(c/\omega_{\rm pe})/ \Delta=26.67$, while
"EM wave" inertial length (this is not an accepted terminology),
$c/\omega_0$, is resolved with 40 grid points, i.e.
$(c/\omega_0)/ \Delta=40$. This means that the numerical resolution used 
is appropriate to resolve the smallest relevant electron kinetic scales
considered. 
In all our time-distance plots in
figures \ref{fig1}-\ref{fig6} we use 500 time snapshots,
which means that data is stored every
$\Delta t=0.2/\omega_0$. This ensures smoothness of
the time-distance plots. 

\section{The results}

In is section we present the results of our numerical simulations 
detailed in table \ref{t1}.
Below we also provide the justification for the number density 
expressions used in this work.

The electron and ion number densities are set 
using EPOCH's  conditional function called 
if($a,b,c$). 
If $a$ condition is {\it true}, the function returns $b$, otherwise the function returns $c$. For example,
in the case 1b or 1d we set electron and ion number densities using
\begin{eqnarray}
n_{\rm e,i}(x)&=& {\rm if~[(x<512/n_x x_{max})}  \nonumber \\ 
&{\rm or}&{\rm (x>1024/n_xx_{max}),10^{-2}n0,n0]},
\label{eq5}
\end{eqnarray}
which means that this conditional function returns the following
expression
\begin{equation}
n_{\rm e,i}=
\begin{cases}
   10^{-2}n_0,&\text{if }(x<512/n_x x_{\rm max})\text{ or }(x>1024/n_x x_{\rm max})\\
    n_0, & \text{otherwise}.
\end{cases}
\label{eq6}
\end{equation}
Here the grid points number
512 and 1024 in x-direction correspond to $x=12.8$ and $x=25.6$ respectively.
Note EPOCH maths parser uses $lt$ instead of $<$ and $gt$ instead of $>$ signs,
see EPOCH user manual section 3.19 for details.
In cases 1b or 1d
the number density (given by equation(\ref{eq5}))
is similar to  case 2 (figure \ref{fig2}(a)), but in 
case 1b or 1d has $n_y=3$, while for case 2
we set $n_y=128$. Number density at time zero
is plotted in case 2 only (figure \ref{fig2}(a)). 
Essentially this is a slab of plasma with top hat density of 
$n_0=10^{15}$ m$^{-3}$ between 512 and 1024 grid points in $x$-direction,
while on domain edges number density drops to $10^{-2}n_0$.
For the case 2 we use number density given by equation (\ref{eq5}).
For the cases 3 and 4 the number density used is 
given by equation (\ref{eq7})
\begin{eqnarray}
n_{\rm e,i}(x)&=& {\rm if~[(x<512/n_xx_{\rm max})}  \nonumber \\ 
&{\rm or}&{\rm (x>1024/n_x x_{\rm max}),10^{-2}n0,n0]}\nonumber \\ 
&-& {\rm F \times if~[(x>896/n_x x_{\rm max})}  \nonumber \\ 
&{\rm or}&{\rm (x<640/n_x x_{\rm max}),0,n0]}.
\label{eq7}
\end{eqnarray}
This means that we subtract ${\rm F} \times n_0$ in the region 
between 640 and 896  grid points in $x$-direction.
Quantity $1-{\rm F} \times n_0$ is the plasma minimal number 
density in the slab.
Note that for case 3, ${\rm F}=0.2$ and for case 4, ${\rm F}=0.99$, 
respectively.
An explanation for the factor $\rm F$ will be given below
after the EM wave driving will be specified.
Number densities for the cases 3 and 4 graphically
are shown in figures \ref{fig3}(a) and \ref{fig4}(a), respectively.

For the cases 5 and 6 the number density used is
given by equation(\ref{eq8})
\begin{eqnarray}
n_{\rm e,i}(x,y)&=& {\rm if~[(x<512/n_x x_{\rm max})}  \nonumber \\ 
&{\rm or}&{\rm (x>1024/n_x x_{\rm max}),10^{-2}n0,n0]}  \nonumber \\ 
&\times & \Biggl( 1- {\rm F} \times \biggl[e^{-{(y-24/n_y y_{\rm max})^8}/{L_y^8}}+
e^{-{(y-64/n_y y_{\rm max})^8}/{L_y^8}}\nonumber \\
&+&e^{-{(y-104/n_y y_{\rm max})^8}/{L_y^8}}\biggr]
\times e^{-{(x-768/n_x x_{\rm max})^8}/{L_x^8}}\Biggr),
\label{eq8}
\end{eqnarray}
where $L_y=y_{\rm max}/10$ and $L_x=x_{\rm max}/64$
are the width of density structures and $y$- and $x$-directions.
Number densities for the cases 5 and 6 are plotted
in figures \ref{fig5}(a) and \ref{fig6}(a), respectively. 
In equation (\ref{eq8}) for case 5 we set ${\rm F}=0.2$ 
and for case 6 it is set to ${\rm F}=0.99$, 
respectively.
Note that 768th grid is right in the middle 
between 512 and 1024 grid points in $x$-direction.
This is the interval  where both plasma and
metamaterial are present in the cases 5 and 6.

\begin{figure*}[!htb]
\centerline{\includegraphics[width=0.7\textwidth]{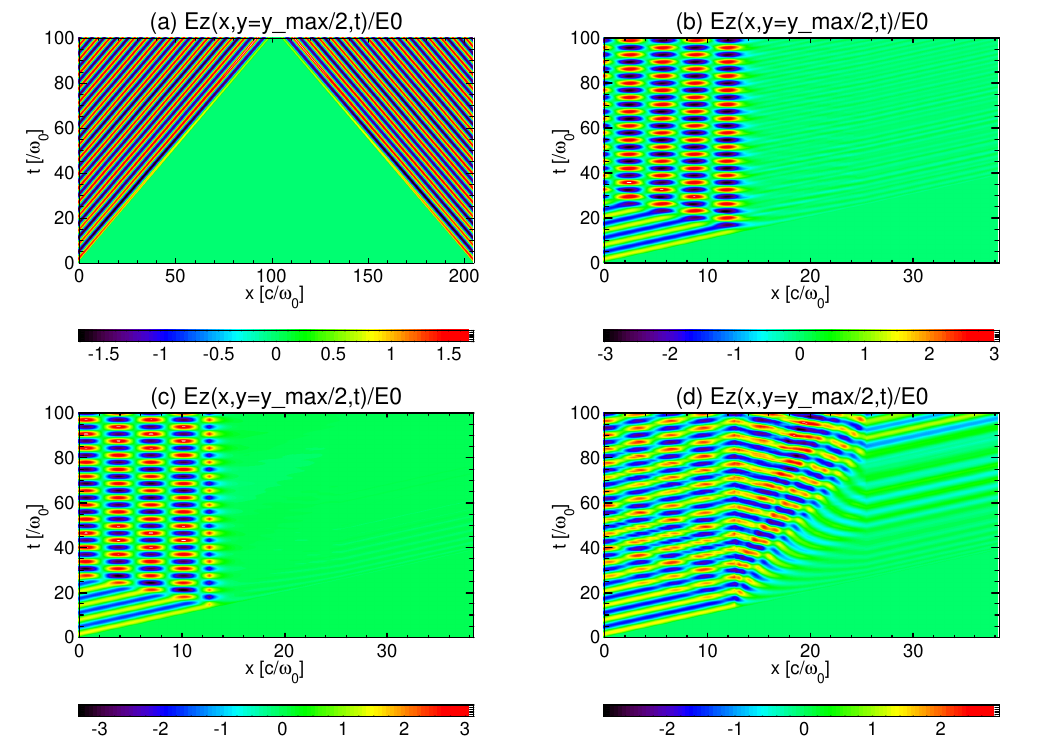}}
\caption{Time-distance plot of  $E_z(x,y=y_{\rm max}/2,t)/E_0$.
(a) case 1a, (b) case 1b, (c) case 1c and (d) case 1d. See table \ref{t1}
for details.}
\label{fig1}
\end{figure*}

\begin{figure*}[!htb]
\centerline{\includegraphics[width=0.7\textwidth]{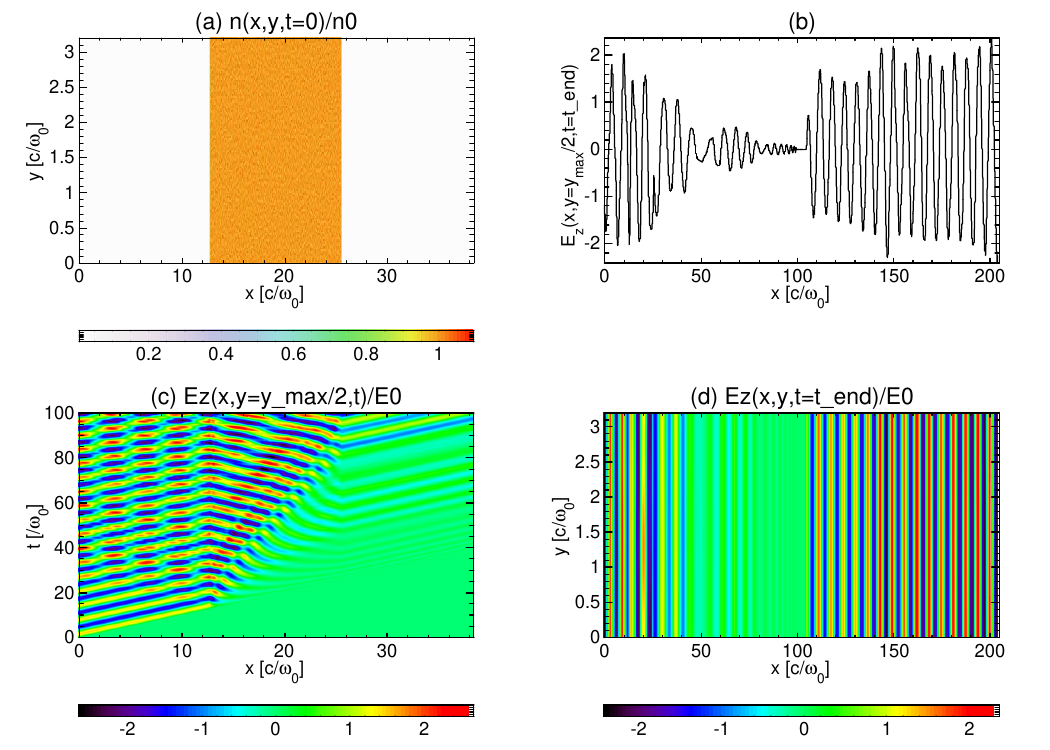}}
\caption{Case 2: (a) $n(x,y,t=0)/n_0$, (b) a line profile of 
$E_z(x,y=y_{\rm max}/2, t=t_{\rm end})$,
(c) time-distance plot of  $E_z(x,y=y_{\rm max}/2,t)/E_0$,
(d) $E_z(x,y,t=t_{\rm end})/E_0$.} 
\label{fig2}
\end{figure*}

\begin{figure*}[!htb]
\centerline{\includegraphics[width=0.7\textwidth]{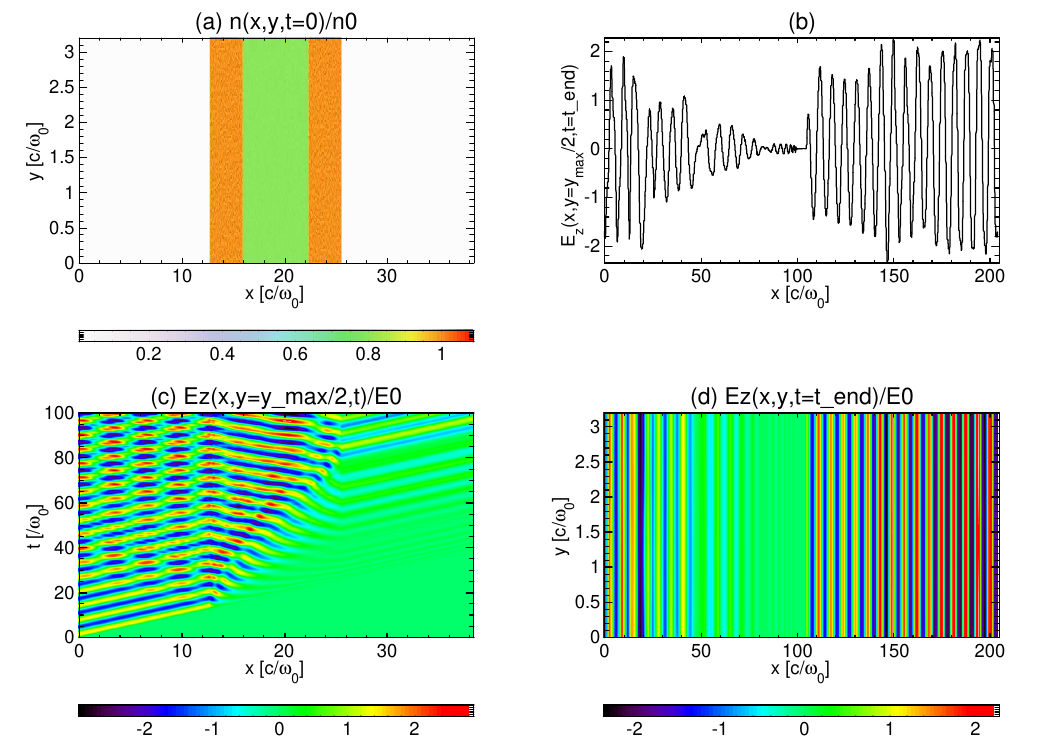}}
\caption{The same as in figure \ref{fig2}, but now for case 3.}
\label{fig3}
\end{figure*}

\begin{figure*}[!htb]
\centerline{\includegraphics[width=0.7\textwidth]{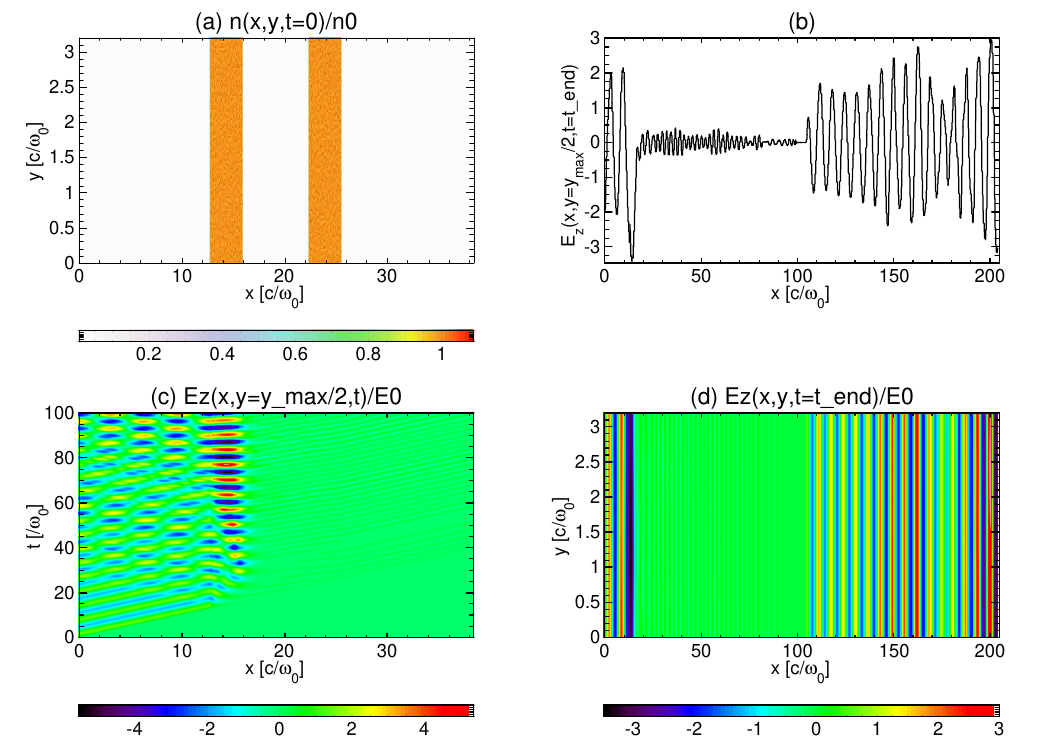}}
\caption{The same as in figure \ref{fig2}, but now for case 4.}
\label{fig4}
\end{figure*}

With number densities in all numerical runs specified, next we
describe how we generate (drive) EM waves.
Similar to \cite{iwai2017} we
drive $x=1 \Delta$ cell as described in Appendix A.
Here we only need to explain the driving field amplitude
$E_0=760199.13$ V m$^{-1}$, which produces linearly
polarized in $z$-direction EM wave that propagates in 
both positive and negative $x$-directions.
Note that because of periodic boundary conditions used,
left propagating wave (i.e one that travels 
in the negative $x$-direction), re-enters the simulation domain
from the right edge of the domain, as shown in figure \ref{fig1}(a).
We also need to comment on the frequency $\omega_0=1.1893397\times 10^9$
Hz radian.
Firstly, $E_0$ is set to
$E_0=0.25 \omega_{\rm pe} m_e c /q_e=760199.13$ V m$^{-1}$.
The relevant electric field scale in this context is
so called wave breaking electric field.
The relevant parameter is 
$a=q_e E_0/(m_e \omega_0 c)$ \cite{esarey2009}.
When $a\geq 1$, the electron quiver motion is relativistic and the
EM-plasma interaction is nonlinear (wave breaks, i.e.
over-turns due to non-linearity). 
EM wave stays linear otherwise for $a\leq 1$. For our value of $E_0$ used
$a=0.375$, which means our results stay
in the moderately linear regime.
Secondly, we set $\omega_0=\omega_{\rm pe}/1.5=1.1893397\times 10^9$ Hz rad.
This means that EM wave cannot propagate through plasma
having number density $n_0=10^{15}$ particles per m$^{-3}$, as $\omega_0$ is
is a factor of 1.5 below the plasma frequency.

We now can explain the use of ${\rm F}=0.2$ and ${\rm F}=0.99$ factors in equations
(\ref{eq7}) and (\ref{eq8}) (see also table \ref{t1}).
$(1-{\rm F})\times 1.5=1.2$ for ${\rm F}=0.2$ and 
$(1-{\rm F})\times 1.5=0.015$ for ${\rm F}=0.99$.
In the first case ($(1-{\rm F})\times 1.5=1.2$) plasma metamaterial composite
is {\it transparent} to EM wave with negative phase speed
propagation, while in the second case ($(1-{\rm F})\times 1.5=0.015$) plasma metamaterial 
composite is {\it opaque} to EM wave.


\subsection{Quasi 1D Results}

One of the aims of this work is to extend previous one dimensional
studies of EM wave propagation on an over-dense 
plasma-metamaterial \cite{iwai2017} 
composite into two spatial dimensions.
This is achieved in 4 numerical runs for cases 1a, 1b, 1c and 1d, see table \ref{t1} for details.
The cases 1(a) to 1(d) are 
considered in order to validate EPOCH 2D 
by comparing its results with Ref. \cite{iwai2017}. 
EPOCH 2D was modified as explained in Appendix A.
The numerical grid is essentially one dimensional:
$n_x\times n_y=8192 \times 3$. This is because EPOCH 2D needs at least 3 grid
points in $y$-direction. 8192 number of grids in $x$-direction was
chosen so that left- and right-propagating EM waves, due to periodic boundary
conditions,
do not collide by the end simulation time of $t_{\rm end}=100/\omega_{\rm pe}$.
This is clearly seen in figure \ref{fig1}(a), where we show
time-distance plot of  $E_z(x,y=y_{\rm max}/2,t)/E_0$ when 
electron and ion number densities are set to $10^{-2}n_0$, i.e.
essentially plasma is so {rarefied (well below
under critical density)}, it can be regarded as a vacuum.
Figure \ref{fig1}(a) corresponds to case 1a.
In figure \ref{fig1}(b) we plot the same electric field component
but for the case 1b, i.e. when plasma is present, but metamaterial is not.
Because plasma is overdense, EM wave with frequency 
$\omega_0=\omega_{\rm pe}/1.5$, cannot propagate through plasma and
a standing EM wave forms.
In figure \ref{fig1}(c) we plot the same 
but for the case 1c, i.e. when plasma is absent ($n_{e,i}=10^{-2}n_0$), but metamaterial is present.
Because metamaterial frequency is $\omega_{\rm m}=1.7 \omega_0$, 
the metamaterial $\mu_m$ is negative so again
EM wave with frequency 
$\omega_0$, cannot propagate through plasma-metamaterial region and
a standing EM wave forms. 
In figure \ref{fig1}(d) we plot the same 
but for the case 1d, 
i.e. when both plasma and metamaterial are present and form a composite
between grid points 512 and 1024 in $x$-direction. We now see
that in the plasma metamaterial composite region which 
lies in the interval $12.8 < x < 25.6$
EM wave with frequency 
$\omega_0$ propagates with negative phase speed (and positive group speed)
because
effective permittivity of the plasma $\varepsilon_p$ and 
the effective permeability of the metamaterial $\mu_m$ are both negative.
We note that our 
figure \ref{fig1} matches closely figure 2 from Ref.\cite{iwai2017}.
Since our simulation parameters for case 1 
($\omega_{\rm pe}=1.5 \omega_0$ which corresponds to
$\varepsilon_p = -1.25$ and $\omega_{\rm m}=1.7 \omega_0$
which corresponds to $\mu_m = -1.89$)
are similar, to that of Ref.\cite{iwai2017}
hence the similarly of the time-distance plots 
is successfully achieved. 

In conclusion, 
 figure \ref{fig1} 
is a successful reproduction of 1D results of Ref.\cite{iwai2017},
but now with EPOCH 2D.


\subsection{2D Results}

Next, in order to study 2D effects in the subsequent runs,
for the testing purposes, in case 2 
now the grid size is increased in $y$-direction to 128. That is
in case 2 everything else is the same as in 
case 1d, but grid size is
$n_x\times n_y=8192 \times 128$.
In figure \ref{fig2}(a) we show a number density slab of plasma
that lies in the intervals $12.8 < x < 25.6$ and 
it is uniform in $y$-direction with
$0 < y < 3.2$. We note that $n(x,y,t=0)/n_0$ in these intervals
is close to unity, i.e. $n(x,y,t=0)/n_0=1$ -- see color table of 
figure \ref{fig2}(a). In figure\ref{fig2}(b)
we plot line profile of electric field
$E_z(x,y=y_{\rm max}/2, t=t_{\rm end})$
in the middle grid point in $y$ at the final simulation 
time. Note the different $x$-range  of plots used
in figure\ref{fig2}(a) and figure\ref{fig2}(b).
In the latter the entire simulation domain length is
shown, in the former only the area around plasma-metamaterial
composite is presented.
The purpose of figure\ref{fig2}(b) is to prove that
the electric field of the EM wave that goes in
negative $x$-direction and because of periodic
boundary conditions re-appears on the right-side
of the domain, does not collide with EM wave
that goes in positive $x$-direction.
We consider EM wave that travels in
positive $x$-direction as a main object of our study.
The negative $x$-direction going wave can be regarded
as a parasitic wave that cannot be avoided due to
imperfect means of EM wave excitation.
Essentially figure \ref{fig2}(b) proves that because of
the absence of the left- right-travelling EM wave collision,
any behaviour for $x>100$ can be ignored.
In figure \ref{fig2}(c) we show
time-distance plot of of  $E_z(x,y=y_{\rm max}/2,t)/E_0$.
We note that this plot is very similar (if not identical)
to \ref{fig1}(d). This is because of the uniformness in 
$y$-direction of the number density, whether $n_y=3$ or 128
it makes no difference.
The same conclusion follow from figure
\ref{fig2}(c) that in the region of plasma 
metamaterial is present ($12.8 < x < 25.6$)
we witness negative phase speed (and positive group speed)
EM wave propagation in an over-dense plasma.
In figure \ref{fig2}(d) we plot
$E_z(x,y,t=t_{\rm end})/E_0$. The purpose of this plot
is to ascertain that no $y$-variation is seen as
the problem is uniform in $y$-direction.
Indeed, we see from figure \ref{fig2}(d) that this is the case.

Cases 3 and 4 aim to study 
blocking of EM waves by 
plasma-metamaterial composite barriers.
Thus, in case 3 we consider a shallow slab (a barrier)
where plasma number density is dropping to $0.8 n_0$,
while in case 4, number density is dropping to $0.015 n_0$,
both in the interval 
between 640 and 896  grid points in $x$-direction.
In the normalized units this is
$12.8\times640/512 <x<12.8\times896/512$
i.e. $16 <x<22.4$, which can be seen
as light-green color strip in the middle of  figure
\ref{fig3}(a) and a white color strip in the middle of  figure
\ref{fig4}(a).
In case 3 ($(1-{\rm F})\times 1.5=1.2$) plasma metamaterial composite
should be transparent to EM wave with negative phase speed
propagation, while in case 4 ($(1-{\rm F})\times 1.5=0.015$) 
plasma metamaterial 
composite should be  opaque to the EM wave.
That is exactly what we see in figures \ref{fig3}(c) and \ref{fig4}(c):
in figure \ref{fig3}(c) the EM wave propagates freely
though the entire plasma metamaterial composite ($12.8<x<25.6$),
while in figure \ref{fig4}(c) EM wave is blocked
by the deep density drop $16 <x<22.4$, hence a standing
EM wave forms near the $x=16$ edge.
This way, we achieve blocking of EM wave by dropping
density in the plasma metamaterial composite to an appropriate value,
such that $(1-{\rm F})\times 1.5< 1$ condition is met.
In figure \ref{fig3}(b) we see similar behaviour of electric field as in
\ref{fig2}(b), but figure \ref{fig4}(b) is significantly
different, we gather that 
$E_z(x,y=y_{\rm max}/2, t=t_{\rm end})$ amplitude
is significantly reduced and the frequency of the
EM wave is increased. 
We conjecture that only non-linear harmonics can
tunnel through the density drop which acts as
a barrier to negative phase speed propagation
seen in figures \ref{fig2}(b,c) figure \ref{fig3}(b,c).
For consistency we also check spatial structure
of EM wave $E_z(x,y,t=t_{\rm end})/E_0$ in figures
\ref{fig3}(d) and \ref{fig4}(d).
Indeed we see in figure \ref{fig4}(d) that EM wave that propagates 
in the positive $x$-direction, does not move
beyond $x>16$, as it is effectively blocked.

\begin{figure*}[!htb]
\centerline{\includegraphics[width=0.7\textwidth]{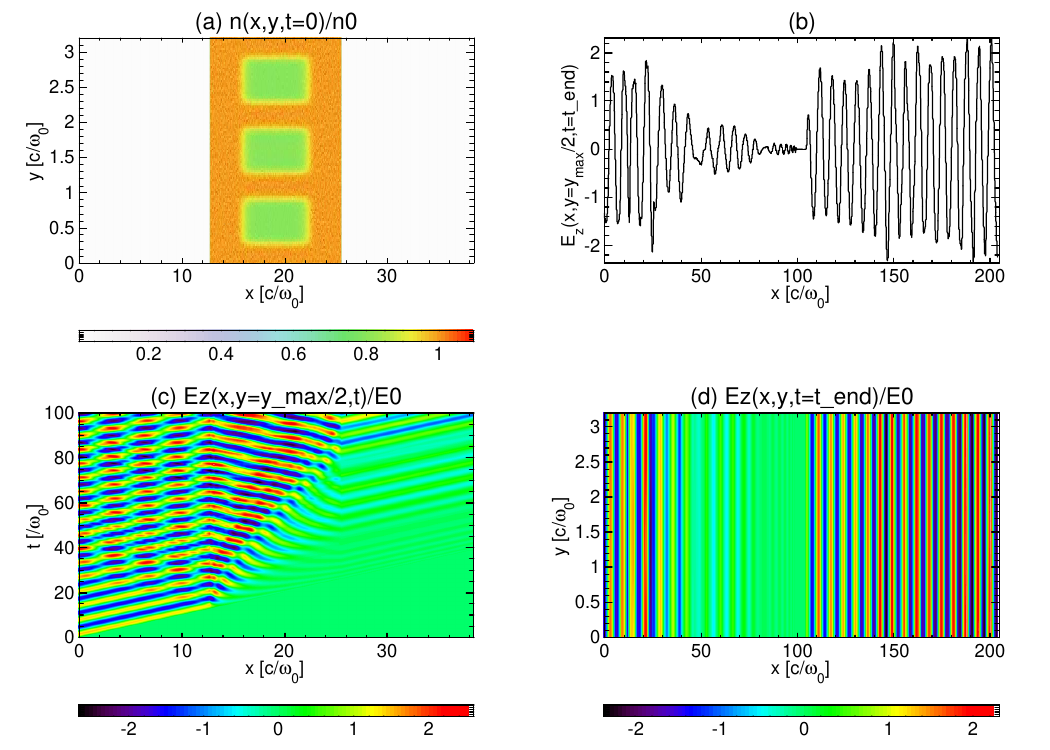}}
\caption{The same as in figure \ref{fig2}, but now for case 5.}
\label{fig5}
\end{figure*}

\begin{figure*}[!htb]
\centerline{\includegraphics[width=0.7\textwidth]{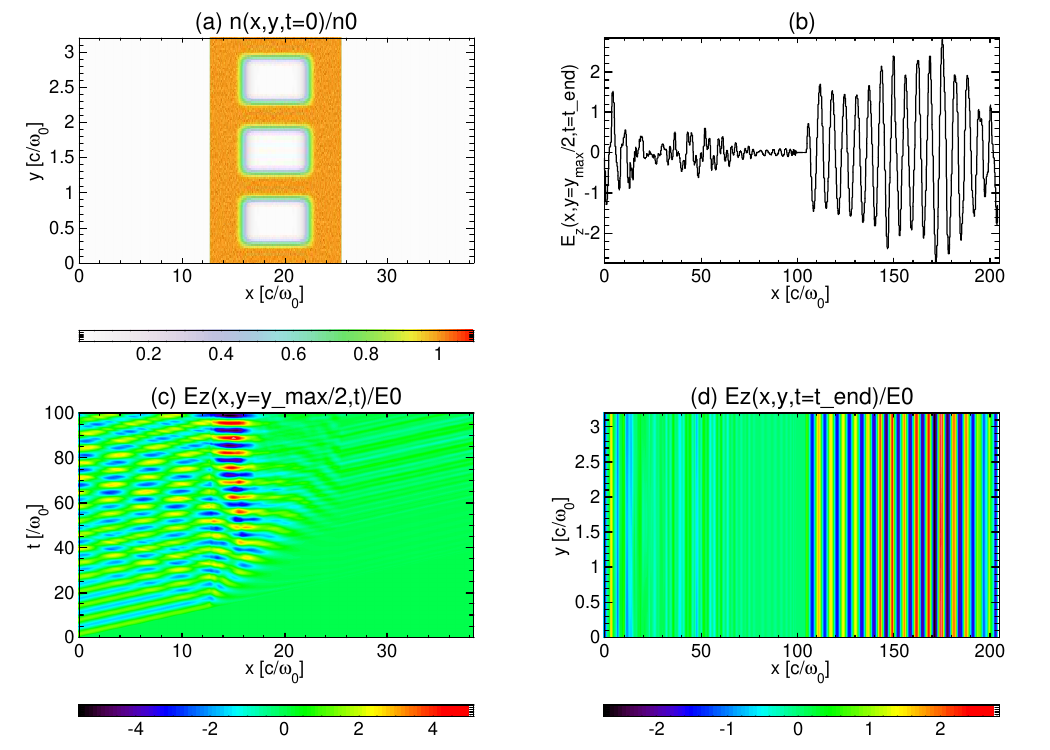}}
\caption{The same as in figure \ref{fig2}, but now for case 6.}
\label{fig6}
\end{figure*}

Cases 5 and 6 also aim to study 
blocking of EM waves by 
plasma-metamaterial composite, but instead of a slab we
consider 
two-dimensional density rectangular 
depletions (DRD) that can be seen
in figures \ref{fig5}(a) and \ref{fig6}(a),
as light-green and white color boxes.
In them, the number density drops to
 $0.8 n_0$ in case 5 (as in case 3)
and $0.015 n_0$ in case 6 (as in case 4).
Their dimensions are
$x_{\rm rect}=22-16=6$ and
$y_{\rm rect}=0.64\approx1$ i.e. $y_{\rm rect}\approx c/\omega_0$.
The gaps between DRDs are 0.32, i.e. $\approx (1/3)\times (c/\omega_0)$.
Essentially, we wanted to add some structuring
in $y$-direction with structures on the scale of 
EM wave inertial length of $\approx c/\omega_0$.
The results of the last two entries
from table \ref{t1} 
are shown in figures \ref{fig5} and \ref{fig6}.
We gather from  figures \ref{fig5} and \ref{fig6}
that shallow DRDs are transparent to
EM wave, while deep ones block EM wave, as expected.
Next, we focus our attention on the differences
from cases 3 and 4, presented earlier
in figures \ref{fig3} and \ref{fig4}.
It can be seen in figure \ref{fig6}(b)
that electric field from has complex
beating pattern (multiple harmonic beating). This is 
in contrast to figure \ref{fig4}(b)
where high frequency harmonics show a regular
spatial pattern (no beats).
Figure \ref{fig6}(d) does not show
any structuring in $y$-direction,
which seams somewhat counter intuitive.
As, for example, it seems not clear
at first sight, 
why in-between DRDs across $y$-coordinate 
where  normalized plasma number density is
unity, i.e. in the four light-brown color strips
on top and bottom  of
white rectangles in figure \ref{fig6}(a),
no clear EM wave propagation seen.
Our explanation is that shortness
of the considered gaps,
$\approx (1/3) \times (c/\omega_0)$,
is not sufficient for apparent EM wave front 
propagation as the wave interacts
with plasma on circa $> c/\omega_0$ scale.
Nonetheless, the beats are seen in \ref{fig6}(b),
which means there is some propagation
in the gaps, hence the beat pattern.
Ideally, in retrospect, we surmise that
a bigger than $n_y=128$ could have been considered.
Nonetheless, we still think the considered
$y$-coordinate structuring in cases 5 and 6
provide a useful information.
Future, a more elaborate study, could
potentially consider different shapes of DRDs and 
different separation distances, which could
be a subject to next work(s).

In conclusion,
here we established that plasma-metamaterial composite 
can block EM waves by both slab and DRD configurations.
This happens by forming a standing wave that subsequently damps
at the edge of an opaque region.

\section{Conclusions}

The aim of this work is two-fold: (i) to extend previous 1D studies
\cite{iwai2017} of electromagnetic (EM) waves propagation in an over-dense plasma-metamaterial composite into 2D spatial dimensions
 and (ii) to study blocking of EM waves by the
composite 2D structures that act like barriers, which 
is analogous to EM wave
trapping by preformed density cavities 
in near-critical density plasmas \cite{luan2013}.
Our study comprises of several logical
steps:\\
1. In Appendix A we presented 
a numerical recipe how to include the presence of
metamaterial in plasma by 
adding additional current $\mathrm{J_m}$
using EPOCH 2D PIC code.\\
2. We then validated the modification of
EPOCH code (figure \ref{fig1}, cases 1a, 1b, 1c and 1d, see table
\ref{t1}) 
by a successful reproduction of 1D results of Ref.\cite{iwai2017}
with EPOCH 2D on a $n_x\times n_y =8192\times 3$ grid.\\
3. With a purpose to study 2D effects in this work,
in case 2 we increased grid size $y$-direction to 128
and ascertained negative phase speed propagation
(figure \ref{fig2}(c)) which is nearly identical to 
figure \ref{fig1}(d) (the quasi-1D case with $n_y=3$ grids along $y$-axis).\\
4. Next, we wanted to study EM wave blocking
by a slab of density depletion (cases 3 and 4).
Thus, we set plasma number density according to equation (\ref{eq7}).
We find that in case 3 ($(1-{\rm F})\times 1.5=1.2$) plasma metamaterial composite is transparent to EM wave. In this case EM
wave propagates with a negative phase speed,
 while in case 4 ($(1-{\rm F})\times 1.5=0.015$) plasma metamaterial 
composite is opaque to the EM wave. Thus,
forming a standing wave at the edge 
of an opaque region.
Here the generic conclusions are as following: (i) if
$(1-{\rm F})\times n_0$ is the minimal number density 
of the plasma slab
and 
(ii) $\rm f=\omega_{\rm pe}/\omega_0$ is the factor by
which EM wave's angular frequency $\omega_0$ is less than electron
plasma frequency $\omega_{\rm pe}$, {\it then}
the condition for EM wave blocking by the slab is 
\begin{equation}
(1-{\rm F})\times f < 1.
\label{eq9}
\end{equation}
We also find that when condition given 
by equation (\ref{eq9}) is met, EM  amplitude
is significantly reduced and the frequency of the
EM wave is increased (figure \ref{fig4}(b)). 
This leads to a conclusion that only non-linear harmonics can
tunnel through the density drop which acts as
a barrier to negative phase speed propagation
found in 
figures \ref{fig2}(b,c) figure \ref{fig3}(b,c).\\
5. Further, we aimed to study EM wave blocking
by two-dimensional DRDs (cases 5 and 6).
Therefore plasma number density was set
according to equation (\ref{eq8}).
We find that for the sizes and shapes of
DRDs considered, behaviour is similar to the 
plasma slab density depletion.
The main difference is
that electric field from. It now has complex
beating pattern (figure \ref{fig6}(b)). 
Such behaviour is 
in contrast to figure \ref{fig4}(b), where
 where high frequency harmonics show a regular
spatial pattern  and no beats are present.
For two-dimensional DRDs the EM wave blocking condition
is also given by equation (\ref{eq9}).

In summary,  the barriers for EM wave propagation 
are created when metamaterial spatially co-exists with a plasma 
density depletion in a form of a slab or two-dimensional 
DRDs. 
Such approach is analogous  to EM
wave trapping by preformed density cavities in near-critical
density plasmas investigated by Ref. \cite{luan2013}. 
We have shown that plasma metamaterial composite allows 
to block EM waves by both a slab (figure \ref{fig4}(c)) 
and DRD configurations (figure \ref{fig6}(c)).
In both cases  a standing wave at
the edge of an opaque region is formed. 
The EM wave blocking condition is established by
equation (\ref{eq9}), which is supported by our 
PIC numerical simulations.
In our approach metamaterial {\it acts like an agent that
lets EM wave to enter over-dense plasma}.
The subsequent density drop {\it makes plasma opaque
to the EM wave},  effectively blocking it in form
of a standing wave at the edge of the barrier
(density depletion slab or two dimensional DRDs).

Ref. \cite{luan2013} established that in a situation when a
 near-critical density plasma contains a preformed
density cavity, EM wave is trapped and becomes a standing wave. 
Their 2D PIC simulations have shown that
within $100-200/\omega_0$ the standing wave damps away.
We only show numerical results with end simulation time
of $100/\omega_0$, but confirm similar result to  
Ref. \cite{luan2013} that within few hundred 
$1/\omega_0$ EM wave indeed damps.
We suggest that the wave damping can offer
applications such as heat deposition or 
substrate materials (micro)machining \cite{stockdale1999} 
depending on EM wave intensity.
Low intensity EM waves would simply heat the plasma,
potentially, high intensity EM waves can be used 
for substrate materials (micro)machining.
From figure \ref{fig4}(c) (density depletion slab case)
and figure \ref{fig6}(c) (two-dimensional DRDs case)
we see that length scale of the standing wave
pattern is $(1-2) c/\omega_0 =0.25-0.5$ m.
Obviously this is the value for 
$n_0=10^{15}$ particles per m$^{-3}$ used, because
number density sets $\omega_{\rm pe}=1.784\times 10^9$ Hz rad,
which, in turn, sets
$\omega_0=\omega_{\rm pe}/1.5=1.189\times 10^9$ Hz rad.
Setting higher number density  say 
$n_0=10^{19}$ m$^{-3}$, can make $c/\omega_0$
as small as 2.5 mm.

{We would like to stress the difference in approach
by \citet{luan2013} and this work. 
In particular, \citet{luan2013} consider situation
when 
the strength of the electric field is high enough, electron density 
increases such that EM waves tends to be reflected by the plasma layer. 
Hence, in that case, the plasma is first allowing propagation of high-power 
EM pulse, and then blocking it as a function of time when electron 
density increases. In our case, the background number density 
does not change markedly as the time progresses.
We do not keep it artificially static, it slowly evolves in
time as PIC simulation progresses. Thus our a
potential applications are different from that of \citet{luan2013}.
Our results potentially  suggest yet unexplored and/or unknown 
 applications of controlling EM waves with plasma and metamaterial
 composite. These may include: a more efficient PVD
   (because the reaction rate is proportional to the
product of number densities of the reactant species), 
controlling EM wave propagation 
(EM wave blocking) in invisibility cloaks and 
perhaps something yet to be discovered in the future.}

{We would like to remark that kinetic
PIC electromagnetic simulations are more realistic than fluid FDTD 
simulations to capture the nonlinear effects in the plasma due to 
 incident high-power EM energy. In this paper, we only 
consider low-power cases ($a = 0.375$) where the physics remains linear. It 
 should be interesting to study novel 
features provided by PIC simulations in the {\it nonlinear}
 case, for instance by 
 showing the electron energy distribution modifications such as in 
 \citet{iwai2017}. This will be subject of a subsequent work.}
 
{In the future, it would be also interesting to explore
the novelties
brought about  by the 2D model, by studying, for example, the effect of 
{\it the angle of incidence} of the EM wave on the composite  of plasma  and 
metamaterial slab.}
 
{Yet another future avenue of research could be 
developing an 
analytical model to 
 validate the numerical results. Indeed, air can be modeled as a 
 propagating {\it transmission line}, while 
 plasma as an evanescent one, and 
 finally the combination of plasma with MNG material as a propagating 
 transmission line. It would be feasible to analytically calculate
  their characteristic (i.e., intrinsic) 
 impedance and propagation constant. 
 The valuable end-product of such an analytical calculation could be
  reflection and 
 transmission coefficients of the complex structure
 so that numerical results can be validated by such a model.
Another possibility for an analytical model is to consider
so-called Zakharov's equations in order to describe the
electronic plasma waves. In such model 
the slowly varying envelope of the electric
field is coupled to the low-frequency variation of the 
density of the ions. \citet{colin2004} used it in
a situation when  the laser enters
the plasma, then part of it is backscattered through 
a Raman-type process and a
Brillouin-type process. The Raman and Brillouin 
parts and the laser combine
together to create an electron-plasma wave. 
These four waves interact in order
to create a low-frequency variation of density 
of the ions which has itself an
influence on the four preceding waves.}
 
There are many examples when there is a good
reason to conduct multi-scale numerical simulations
combining ab initio/density functional theory/molecular dynamics-type
simulations \cite{jenkins2000} and meso-scale, particle-in-cell-based 
modelling, such as: 
(i) correlated simulation studies concerning plasma charging and its effects over ten orders of
magnitude in length scales at the lunar terminator
\cite{wang2023},
(ii) dynamic coupling between particle-in-cell and atomistic simulations
to simulate metal surface response to high electric fields \cite{veske2020},
(iii) molecular dynamics model and particle-in-cell model to investigate the physics of ionic electrospray propulsion over 9 orders of 
magnitude in length scale \cite{asher2022},
(iv) multi-scale/hybrid modelling of low temperature plasmas for fundamental investigations and equipment design \cite{kushner2009},
(v) gas discharge plasmas used for various materials
science applications \cite{bogaerts2006}. Future extensions of this work 
will be combining molecular dynamics and particle-in-cell modeling as e.g. in Ref.\cite{veske2020}.

An interesting observation can be also made as follows:
When the de-Broglie wavelength of plasma particles becomes 
of the order of mean inter-particle distance,  quantum effects 
such as  the combined effect of
Bohm potential and exchange 
correlation potential significantly 
modifies the dispersion properties
of e.g. lower hybrid wave \cite{rimza2017}.
In the case when plasma density is
high the exchange correlation potential 
can in principle dominate over the particle dispersion effects 
\cite{maroof2016,jamil2015,shahmansouri2016}.
Broadly speaking 
this exchange correlation potential can be regarded
as a form of the density fluctuation
theory (DFT) \cite{maroof2016}. The influence of exchange 
potential and Bohm force on
the propagation characteristics of 
different type of waves in plasmas 
have been investigated before \cite{maroof2016,jamil2015,shahmansouri2016}.
Since the number densities considered here, in principle,
are high enough to invoke 
the quantum effects such as (i) the exchange correlation potential, (ii) Bohm potential, 
any future works on the subject 
should consider incorporating these for completeness.

\section*{Acknowledgements}
The author would like to thank: 
(i) The two anonymous referees for useful
comments that improved this manuscript and
(ii) The Senior Editor Dr Francesco Taccogna for 
organizing useful and rigorous review process.
\vskip 0.25cm
%%%%%%%%%%%%%%%%%%%%%%%%%%%%%%%%%%%%%%%%%%%%%%%%%%
{\bf Data availability statement.}
The data and numerical codes 
that support the findings of this study are available
from the corresponding author upon reasonable request.
%\vskip 0.25cm
%{\bf Declarations}
%\vskip 0.25cm
%{\bf Author contributions.} The authors contributed equally
%to this work.

\appendix
\section{Appendixes}

Here we present a numerical recipe how to include the presence of
metamaterial in plasma by adding the additional current $\mathrm{J_m}$
using EPOCH PIC code \cite{arber2015}.
The latter can be downloaded from
GitHub repository
\href{https://epochpic.github.io/quickstart.html}{https://epochpic.github.io/quickstart.html}.\\
Step 1: edit src/epoch2d.F90 and after line
 \begin{verbatim}CALL open_files      ! setup.f90\end{verbatim}
 initialize  $\mathrm{J_{m,y}}$ as zero value
\begin{verbatim}jmy(:,:)=0.0_num.\end{verbatim}
Note that "underscore num" is making sure that EPOCH is using F90 double precision, which
is EPOCH's default accuracy. Historically using the compiler auto-promotion of REAL to DOUBLE PRECISION was unreliable, so EPOCH
uses "kind" tags to specify the precision of the code. See section 1.5 in the
EPOCH user manual.\\
Step 2: edit \begin{verbatim}src/shared_data.F90 \end{verbatim}
and in the following line add $\mathrm{J_{m,y}}$ declaration (last entry):
\begin{verbatim}REAL(num), ALLOCATABLE, DIMENSION(:,:) :: & 
ex,ey,ez,bx,by,bz,jx,jy,jz,jmy \end{verbatim} 
Step 3: edit \begin{verbatim}src/housekeeping/mpi_routines.F90 \end{verbatim}
and after  line 388 add
\begin{verbatim}ALLOCATE(jmy(1-jng:nx+jng, 1-jng:ny+jng)) \end{verbatim} 
to allocate $\mathrm{J_{m,y}}$ array.\\
Step 4: edit src/fields.f90
and after  line "CONTAINS" add
two following subroutines in the case of both plasma and metamaterial 
composite are present. 
If metamaterial is absent and only electric field of EM
wave needs to be driven, then only leave the 1st subroutine. 
Below are the said two subroutines:
\begin{verbatim}
SUBROUTINE drive_field(time) 
REAL(num), INTENT(IN) :: time
IF (x_min_boundary) THEN
 ez(1,:) =ez(1,:)+760199.13_num*&
 sin(1.1893397e+09_num*time)
END IF
END SUBROUTINE drive_field

SUBROUTINE drive_mfield(time)
REAL(num), INTENT(IN) :: time
IF (x_min_boundary) THEN
jmy(512:1024,:)=jmy(512:1024,:)+hdt*&
(1.7_num*1.1893397e+09_num)**2*by(512:1024,:)
by(512:1024,:) =by(512:1024,:)-&
hdt*jmy(512:1024,:)
END IF
END SUBROUTINE drive_mfield
 \end{verbatim}
Further,
in the end of src/fields.f90 file, after half-time step and full time step 
the following calls to the said routines should be added:
\begin{verbatim}
    CALL update_e_field
CALL drive_field(time+0.5_num*dt)
    CALL efield_bcs
    CALL update_b_field
CALL drive_mfield(time+0.5_num*dt) 
...
    CALL update_b_field
CALL drive_mfield(time+dt)
    CALL bfield_final_bcs
    CALL update_e_field
CALL drive_field(time+dt)
    CALL efield_bcs
\end{verbatim}

Note that there is no need to modify EM solver part in the src/fields.f90 file.
The fact that presence of metamaterial is
implemented by setting $\mathrm{J_m}$ as in the above subroutine
$\rm jmy(512:1024,:)$ puts a {\it restriction} on number of
processors that can be used in $x$-direction.
The grid cells $512:1024$ must fit on the first
processor. Thus, because $8192/1024=8$, we are
restricted to use no more than 
8 processors in $x$-direction. There is no restriction on
number of processors in $y$-direction.
EPOCH code allows simulation domain decomposition
to be set manually. In this work we have used
EPOCH's nprocx=8 and nprocy=1 setting.
%%%%%%%%%%%%%%%%%%%% REFERENCES %%%%%%%%%%%%%%%%%%

% The best way to enter references is to use BibTeX:

\nocite{*}
\bibliography{paper86_revtex}% Produces the bibliography via BibTeX.

\end{document}